\documentclass[aps,prl,reprint,superscriptaddress,showpacs,amsmath,amssymb,longbibliography,lengthcheck]{revtex4-1}

\usepackage[dvipdfmx]{graphicx}
\usepackage{color}
\usepackage{txfonts}

\begin{document}

\title{Novel shape evolution in Sn isotopes from magic numbers 50 to 82}

\author{Tomoaki Togashi}
\affiliation{Center for Nuclear Study, University of Tokyo, Hongo, Bunkyo-ku, Tokyo 113-0033, Japan}
\author{Yusuke Tsunoda}
\affiliation{Center for Nuclear Study, University of Tokyo, Hongo, Bunkyo-ku, Tokyo 113-0033, Japan}
\author{Takaharu Otsuka}
\affiliation{Department of Physics, University of Tokyo, Hongo, Bunkyo-ku, Tokyo 113-0033, Japan}
\affiliation{Center for Nuclear Study, University of Tokyo, Hongo, Bunkyo-ku, Tokyo 113-0033, Japan}
\affiliation{RIKEN Nishina Center, 2-1 Hirosawa, Wako, Saitama 351-0198, Japan}  
\affiliation{Instituut voor Kern- en Stralingsfysica, KU Leuven, B-3001 Leuven, Belgium}
\affiliation{National Superconducting Cyclotron Laboratory,Michigan State University, East Lansing, Michigan 48824, USA}
\author{Noritaka Shimizu}
\affiliation{Center for Nuclear Study, University of Tokyo, Hongo, Bunkyo-ku, Tokyo 113-0033, Japan}
\author{Michio Honma}
\affiliation{Center for Mathematical Sciences, University of Aizu, Ikki-machi, Aizu-Wakamatsu, Fukushima 965-8580, Japan}

\begin{abstract}
\noindent
A novel shape evolution in the Sn isotopes by the state-of-the-art application of the Monte Carlo Shell Model
calculations is presented in a unified way for the $^{100-138}$Sn isotopes.   A large model space consisting of eight single-particle orbits for protons and neutrons is taken with the fixed Hamiltonian and effective charges, 
where protons in the 1$g_{9/2}$ orbital are fully activated.  
While the significant increase of the $B(E2; 0^+_1\rightarrow2^+_1)$ value, seen around $^{110}$Sn as a function of neutron number ($N$), has remained a major puzzle over decades, it is explained as a consequence of the shape evolution driven by proton excitations from the 1$g_{9/2}$ orbital. 
A second-order quantum phase transition is found around $N$=66, connecting the phase of such deformed shapes to the spherical pairing phase.  The shape and shell evolutions are thus described, covering topics from the Gamow-Teller decay of $^{100}$Sn to the enhanced double magicity of $^{132}$Sn.    
\end{abstract}

\pacs{21.60.-n, 23.20.Lv, 27.60.+j}

\maketitle

\begin{figure}[tb]
  \includegraphics[width=8.4cm]{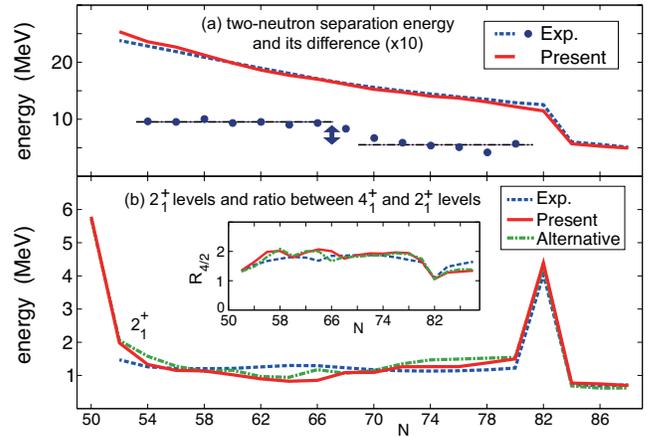}
  \caption{(Color online) (a) Two-neutron separation energy, $S_{2n}(N)$, and (b) 2$^+_1$ levels are shown, 
  as a function of $N$, for experimental  \cite{nudat2} and Present values by blue dashed and red solid lines, 
  respectively.
The inset of the panel (b) displays the $R_{4/2}$ ratio: the 4$^+_1$ level over the 2$^+_1$ level energy.  
In (a),  $\Delta S_{2n} \times 10$ is shown with averages by the dashed-dotted lines.  
In (b), Alternative values are shown by green dashed-dotted lines.
    }
  \label{fig:energy}
\end{figure}

The shape is one of the fundamental concepts in the physics of atomic nuclei, and its variation has been studied from many angles \cite{bohr_mottelson,ring_schuck,casten}.  In such studies, Sn isotopes, where the proton number ($Z$) is equal to the magic number 50, have played an anchor point, as their shapes have been considered to be basically spherical.   Consequently, 
their structure has been often described in the (generalized) seniority scheme \cite{deshalit1963,talmi1971}, where the ground state
is a condensate of a collective pair of two valence neutrons coupled to the angular momentum $J=0$, which can be regarded as a BCS pair with the isotropic amplitudes  \cite{bohr_mottelson,allaart1988,ring_schuck}.
The 2$^+_1$ state is then described as a seniority-two ({\it i.e.}, two-quasiparticle) state, \cite{shlomo1972,vangunsteren1973,ring_schuck}, while rather constant excitation energies are observed for isotopes with even neutron number ($N$) between the magic numbers $N$=50 and 82 \cite{nudat2} (see Fig.~\ref{fig:energy}(b)). 
E2 transition probabilities are direct indicators of the deformation from a sphere to an ellipsoid, and are quantified through $B(E2)$ values \cite{blatt_weisskopf,bohr_mottelson}.  For Sn isotopes, anomalous deviations from the spherical picture were observed for the $B(E2; 0^+_1\rightarrow2^+_1)$ value as a function of $N$
\cite{radford2004,radford2005,banu2005,cederkall2007,vaman2007,doornenbal2008,ekstrom2008,kumar2010,
allmond2011,jungclaus2011,guastalla2013,bader2013,doornenbal2014,allmond2015,umbartzki2016,kumar2017,faestermann2013,pietralla2017,reiter2018}, but   
a unified theoretical description of these anomalies is missing.  In this Letter, we present, for the first time, results of state-of-the-art calculations with the Monte Carlo Shell Model (MCSM) \cite{mcsm_review1,mcsm_ptep} on Sn isotopes, performed in a large model space including single-particle orbits below and above the magic numbers 50 and 82.
We demonstrate that these anomalies, a long-standing challenge to nuclear theory, are now solved, clarifying how the shape and structure evolve in those Sn isotopes.      
 
We perform shell-model (SM) calculations (called also Configuration Interaction calculations) by taking the full $sdg$ Harmonic Oscillator (HO) shell consisting of the 1$g_{9/2,7/2}$, 2$d_{5/2,3/2}$ and 3$s_{1/2}$
single-particle orbits as well as the lower part of the next HO shell, {\it i.e.}, the 1$h_{11/2}$, 2$f_{7/2}$ and 3$p_{3/2}$ orbits.  The same set of the orbits are taken for protons and neutrons, keeping the isospin ($T$) conserved.  
The conventional SM calculation 
is 
not tractable as the maximum 
dimension reaches $7.5\times10^{23}$, beyond $10^{12}$ times its current limit.   
     
\begin{figure}[tb]
\includegraphics[width=8.6cm]{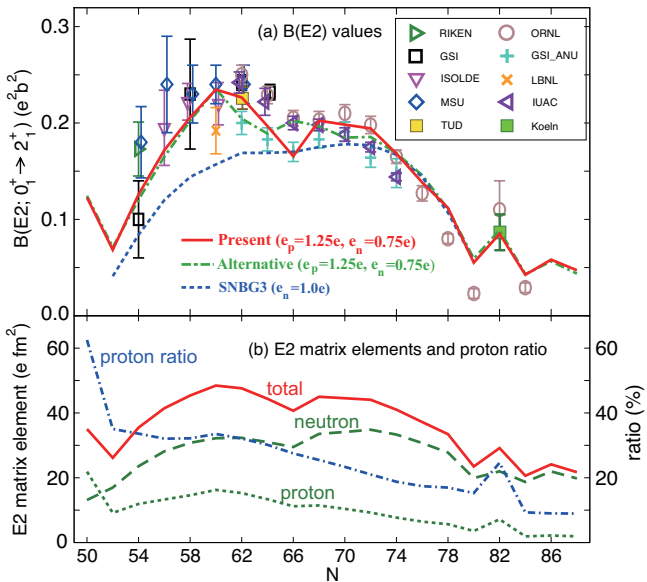}
\caption{ (Color online) (a) Calculated and measured B(E2) values, and (b) E2 matrix elements and proton 
ratio ($\%$).
Experimental data are indicated by symbols shown in the inset with the correspondence: RIKEN   
   \cite{doornenbal2014}, ORNL \cite{radford2004,radford2005,allmond2011,allmond2015}, GSI 
   \cite{banu2005,doornenbal2008,kumar2010,guastalla2013}, GSI\underline{  }ANU \cite{jungclaus2011},  
   ISOLDE \cite{cederkall2007,ekstrom2008}, LBNL \cite{umbartzki2016}, MSU \cite{vaman2007,bader2013},  
   IUAC \cite{kumar2017}, TUD \cite{pietralla2017}, Koeln \cite{reiter2018}.
}
\label{fig:BE2}
\end{figure}

The SM calculations including only valence neutrons in the $N$=50-82 shell
have been carried out 
\cite{banu2005,vaman2007,guastalla2013,back2013,bader2013,faestermann2013,coraggio2015}.   Another example was the SNBG3 Hamiltonian \cite{snbg3}, producing a perfect agreement of the 2$^+_1$ levels with experiment over $N$=52-80 (see Fig.~\ref{fig:energy}(b)).  However, it failed to explain observed bump of $B(E2; 0^+_1\rightarrow2^+_1)$ values around $N$=52-64 (see Fig.~\ref{fig:BE2}(a)).  
Although this objective has been pursued by various theoretical approaches\cite{ansari2005,banu2005,back2013,bader2013,guastalla2013,coraggio2015}, 
no clear picture has been reported\cite{doornenbal2014} and, for instance, an $\alpha$-correlation was discussed instead of the then unsuccessful shell model\cite{vaman2007}.  
It is therefore required to apply the MCSM to Sn isotopes, in order to see if this serious discrepancy can be solved and how.
We mention that Sn isotopes are relevant to the neutrinoless $\beta\beta$ decay \cite{horoi2016} and the processing of the long-lived fission product \cite{trs435}.  

The MCSM can diagonalize the Hamiltonian with many active protons and neutrons in a wide model space
\cite{mcsm_review1,mcsm_ptep}.  
We use a single Hamiltonian throughout this work, aiming at a unified description of the varying structure of the even-$N$ Sn isotopes including proton degrees of freedom.     
The SM Hamiltonian is represented in general by so-called two-body matrix elements (TBME) of the effective nucleon-nucleon interaction between two-nucleon states
where the nucleons are in assigned single-particle orbits and are coupled to given $J$ and $T$.   
In the present work, TBMEs are grouped and were taken from existing, well-tested sets as much as possible, with minor modifications possibly due to different inert cores.  
The TBMEs involving only 1$g_{9/2}$ are taken from the JUN45 set \cite{jun45}.
The SNBG3 set is taken for other orbitals as stated later \cite{snbg3}.    Note that the JUN45 (SNBG3) set was obtained, in the corresponding model space, 
by modifying TBMEs calculated from microscopic interactions called G-matrix \cite{hjensen1995} based on the CD-Bonn \cite{machleidt1989} (N3LO \cite{entem2003}) interaction, so as to better reproduce relevant experimental energies \cite{jun45,snbg3}.  
The $V_{{\rm MU}}$ interaction \cite{vmu} is taken for the rest of TBMEs 
except for some cases stated below.  
It consists of the central part given by a Gaussian 
function in addition to the $\pi$- and $\rho$-meson exchange 
tensor force \cite{vmu}.  The parameters of this Gaussian function were fixed from monopole 
components of known SM interactions \cite{vmu}.
No adjustment is made for the $T=0$ interaction. 
The $T$=1 TBMEs for 1$g_{7/2}$, 2$d_{5/2,3/2}$, 3$s_{1/2}$ and 1$h_{11/2}$ are taken first from the SNBG3 set, and are fine-tuned based on the so-called LC method \cite{honma02,brown06}, so as to reproduce observed 2$^+_1$ and 4$^+_1$ levels of $^{102-138}$Sn.  As the present calculation requires huge computer resources, we calculate only 0$^+_1$, 2$^+_1$ and 4$^+_1$ levels for the fitting purpose.  With the next generation of supercomputers, the situation can be improved.  Thus, levels considered in this fit are quite limited, and the fit here means a minor improvement.  
Regarding the remaining $T$=1 TBMEs, most of them are given by the $V_{{\rm MU}}$ interaction, for which the 
central part is reduced by a factor of 0.75 except for TBMEs involving the $1f_{7/2}$ and $2p_{3/2}$ orbits, similarly to 
\cite{togashi2016}.
On the other hand, $J^{\pi}$=0$^+$ TBMEs are given by the simple pairing ones, being  
$\propto \sqrt{(2j+1)(2j'+1)}$ for orbitals $j$ and $j'$ with appropriately fitted strengths, 
as the $V_{{\rm MU}}$ interaction may not be so suitable.   
All TBMEs are scaled, as usual, in proportion to $A^{-1/3}$ ($A$=$Z+N$).   
The single-particle energies are determined so as to be consistent with the predictions of the 
JUN45 and SNBG3 sets, and those outside them are estimated by a standard Woods-Saxon potential.  
The SM Hamiltonian is thus fixed and kept unchanged, with the results labelled "Present".   For the purpose of comparison, we show the results labelled "Alternative", where the diagonal TBME of $J^{\pi}$=6$^-$ 3$s_{1/2}$-1$h_{11/2}$ state is changed by $\sim$0.1 MeV and 
the $V_{{\rm MU}}$ interaction is used for $T$=1 $J^{\pi}$=0$^+$ TBMEs outside the SNBG3 model.

Figure \ref{fig:energy}(a) shows the two-neutron separation energies, $S_{2n}$, in comparison to experiment
\cite{nudat2}.  The Coulomb correction is included \cite{coulomb1,coulomb2}.  
Note that the fit to the excitation energies leaves an overall common shift for all single-particle energies
open.  It is fixed here to be -0.25 MeV.  The agreement is quite good.

Figure~\ref{fig:energy}(b) shows the 2$^+_1$ levels and, in the inset, the ratio of the 4$^+_1$ level over the 2$^+_1$ level energy, denoted $R_{4/2}$.
The results labeled ``Present'' show the overall trends rather well, with the 2$^+_1$ level and the $R_{4/2}$ ratio both being fairly flat and close to experiment, including certain variations near the magic numbers.  
These agreements suggest that the present Hamiltonian gives a reasonable description of the structure of the Sn isotopes, apart from fine details.  We then look into signatures of 
novel characteristics of Sn isotopes, hidden in the constancy of the energy levels.
      
\begin{figure}[bt]
\includegraphics[width=8.0cm]{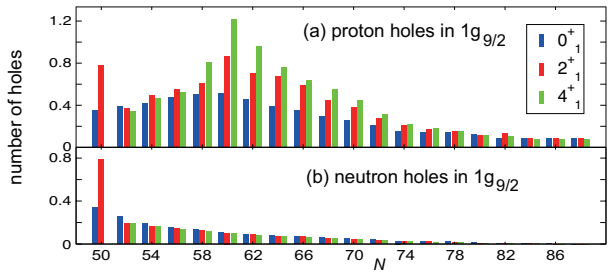}
\caption{ (Color online) 
Occupation numbers of (a) proton and (b) neutron holes in the 1$g_{9/2}$ orbit.
For each value of $N$, histograms for the $0^+_1$(blue),  $2^+_1$(red), $4^+_1$(green) states are shown from left to right, except for the missing $4^+_1$ for $N$=50.
}
\label{fig:occup}
\end{figure}

One of such signatures is the variation of E2 transition strength as $N$ changes.    
As mentioned earlier, Fig.~\ref{fig:BE2}(a) displays $B(E2; 0^+_1\rightarrow2^+_1)$ values, where  
the large bump of experimental values around $N$=60 
\cite{radford2004,radford2005,banu2005,cederkall2007,vaman2007,doornenbal2008,ekstrom2008,kumar2010,
allmond2011,jungclaus2011,guastalla2013,bader2013,doornenbal2014,allmond2015,umbartzki2016,kumar2017} shows distinct discrepancy with theoretical calculations \cite{ansari2005,banu2005,vaman2007,back2013,bader2013,guastalla2013,coraggio2015}.
This discrepancy must be resolved because $B(E2; 0^+_1\rightarrow2^+_1)$ is a sensitive and crucial probe of nuclear shape.  
The present calculation indeed reproduces the measured B(E2) trend quite well, with fixed effective charges, $(e_p, e_n) = (1.25e, 0.75e)$.  
Because of this salient agreement, it is of extreme interest to explore the essential underlying mechanism of  
the structure evolution in the Sn isotopes.  

This mechanism is investigated first in terms of the occupation numbers of the 1$g_{9/2}$ orbit, which is completely occupied in the models such as SNBG3.    
Figure~\ref{fig:occup}(a,b) depict, respectively, the number of proton and neutron {\it holes} in the 1$g_{9/2}$ orbit.
This number is about 0.4 (0.8) for both protons and neutrons in the ground (2$^+_1$) state of $^{100}$Sn, 
yielding a larger B(E2).  As shown in Fig.~\ref{fig:occup}(b), this number for neutrons becomes swiftly smaller as $N$ increases, and almost vanishes for $N>74$.  In contrast, Fig.~\ref{fig:occup}(a) shows that for protons, 
    this number increases up to $N$=60 for all states, and then decreases.   These large numbers of proton holes suggest significant breaking of the $Z$=50 magic core and associated deformation \cite{bader2013}.   Figure~\ref{fig:BE2} (b) displays E2 matrix elements (including effective charges) as well as their decomposition into proton and neutron contributions.  Figure~\ref{fig:BE2}(b) also shows the ratio of the proton contribution in the total matrix element, which exceeds 30\% for $N$=50-64, where the 1$g_{9/2} \rightarrow$ 2$d_{5/2}$ excitation is most important.    
The strong proton excitations and the B(E2) bump are consequences of enhanced effects of the proton-neutron interaction leading to a stronger quadrupole deformation.  These effects and resulting shapes vary as a function of $N$. This proton-neutron interaction, thus crucial, is taken primarily from the $V_{{\rm MU}}$ interaction, which has been fixed \cite{vmu} and well-tested, {\it e.g.} with Zr isotopes \cite{togashi2016}.
We stress that the basic trend presented here is quite robust in this respect.  

The MCSM eigenstate is given by a superposition of MCSM basis vectors. 
Each MCSM basis vector is a deformed Slater determinant, for which intrinsic quadrupole moments $Q_0$ and $Q_2$ can be calculated.  Such $Q_0$ and $Q_2$ can be used as ``partial coordinates'', and a given MCSM basis vector is placed as a circle on the Potential Energy Surface. 
The importance of this basis vector in the eigenstate is expressed by the area of a circle.
This visualization of the shape is called the T-plot \cite{tsunoda2014,otsuka2016}.

\begin{figure*}[tb]
\includegraphics[width=18cm]{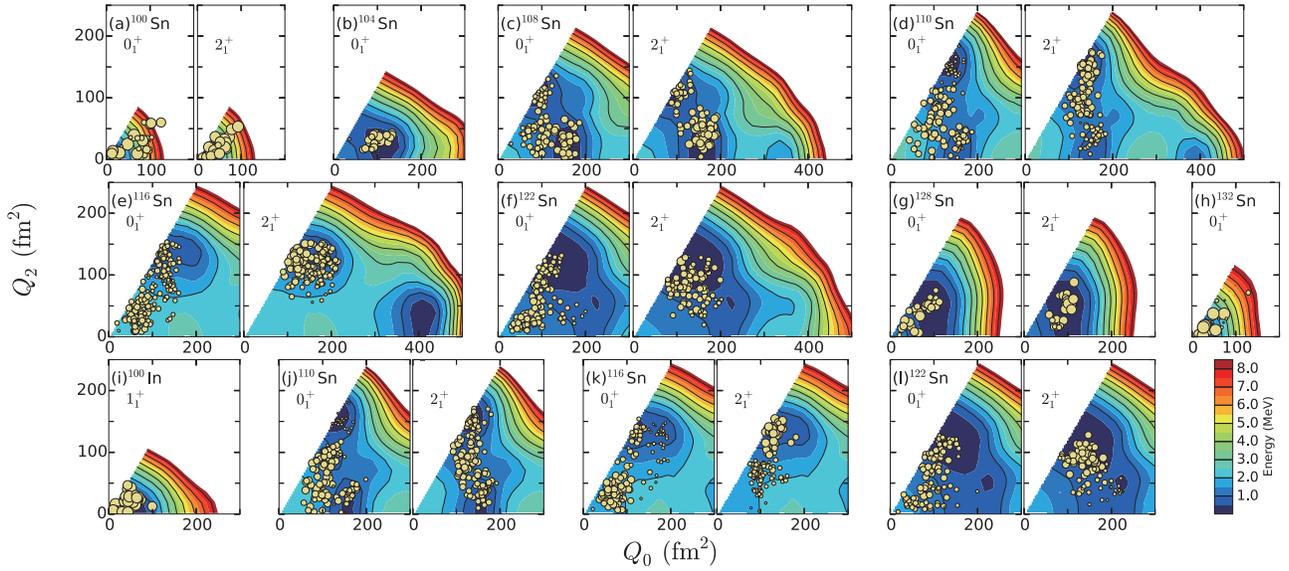}
\caption{ (Color online)
(a)-(h) T-plots for $0^+_1$ and $2^+_1$ states of selected Sn isotopes and (i) T-plot for $1^+_1$ state of $^{100}$In, with the Present interaction. (j)-(l) the same as (a)-(h), with the Alternative interaction. 
}
\label{fig:Tplot}
\end{figure*}

Figure~\ref{fig:Tplot} exhibits the T-plot for selected cases. 
While $^{100}$Sn shows a spherical shape (panel (a)), already in $^{104}$Sn (panel (b)) 
a moderate deformation emerges with no T-plot circles around the spherical limit.   
Such departure from the sphericity is further driven in $^{108}$Sn (panel (c)): the shape is more deformed due to more neutrons, and the tendency towards prolate shape 
arises. 
This is because neutrons occupying the 1$g_{7/2}$ and 2$d_{5/2}$ orbits favor the prolate deformation and likewise protons excited from the 1$g_{9/2}$ orbit can cause the same, producing coherently a prolate shape.  As the 1$g_{7/2}$ and 2$d_{5/2}$ orbits are more filled, the T-plot extends farthest in $^{110}$Sn (panel (d)) with the maximum calculated B(E2) (see Fig.~\ref{fig:BE2} (a)), while the tendency is changed from the prolate to the oblate shape.  Note that a recent B(E2) value on $^{112}$Sn, the next nucleus, \cite{pietralla2017} is in good agreement with the present work.  
For $^{116}$Sn (panel (e)), a notable displacement between the $0^+_1$ and $2^+_1$ T-plot circles appears, which will be discussed later.  
As shown in panel (f), the T-plot circles for the $0^+_1$ state of $^{122}$Sn are shifted from the oblate minimum to the spherical region due to the pairing correlation, whereas those for the $2^+_1$ state are around the minimum, being a quadrupole excitation from the spherical shape.  
In going from $^{122}$Sn to $^{132}$Sn (panels (f,g,h)), the overall spread of T-plot circles is reduced 
gradually, keeping relative ``topological'' relations unchanged to a good extent, as a new feature.
The doubly magic $^{132}$Sn 
exhibits a concentration of T-plot circles near the spherical corner for the ground state (panel (h)).  
We note that the valence mirror symmetry of neutrons is broken severely; {\it e.g.} the T-plot of $^{104}$Sn (panel (b)) is more spread than that of $^{128}$Sn (panel (g)).  

The quadrupole component of the proton-neutron interaction produces stronger deformation as $N$ increases up to $\sim$60.  Beyond this, neutron 1$g_{7/2}$ and 2$d_{5/2}$ orbits are more-than-half filled, which makes the deformation saturated and then weaker.  Beyond $N\sim$66, the pairing correlations take over, and the spherical ground states appear.     
The shell evolution driven by the tensor and central forces \cite{otsuka2005,vmu} contributes: the proton 1$g_{9/2}$- 2$d_{5/2}$ splitting, for instance, becomes wider, first ($N$ up to $\sim$66) by neutrons in the 1$g_{7/2}$ and 2$d_{5/2}$ orbits and later by neutrons in the 1$h_{11/2}$.  Thus, the $Z$=50 gap increases gradually, leading to the highly stable doubly-magic $^{132}$Sn.  
This is not the full story, however.

Figure~\ref{fig:BE2}(b) depicts E2 matrix elements as functions of $N$, with kinks at $N$=66. 
These kinks imply that the dynamical mechanism may change there.  Figure~\ref{fig:energy}(a) shows
$\Delta S_{2n}$=$-(S_{2n}(N)$-$S_{2n}(N-2))$, which is remarkably constant for $N$=54-66 and for $N$=70-80, separately.  Figure~\ref{fig:energy}(a) also shows their averages, including a discontinuity between them.   Since $\Delta S_{2n}$ corresponds to the second derivative of the ground-state energy, this discontinuity points to a second-order quantum phase transition with control parameter $N$ \cite{qpt1,qpt2}.  
While Figure~\ref{fig:energy}(a) shows experimental $\Delta S_{2n}$ values, a similar overall trend is obtained in the present calculation.
Coming back to Fig.~\ref{fig:BE2}(b), the ``derivative'' of E2 matrix elements as a function of $N$ shows discontinuity similarly to $\Delta S_{2n}$.     
This is consistent with the change of the T-plot pattern.  These experimental and theoretical observations imply coherently:  
until $N\sim$66, the moderate deformation phase dominates the low-lying eigenstates, and the transition occurs such that the pairing phase takes over with the seniority-zero
(pair-condensed) ground state and its excitations.
We note that the present case differs from the first-order quantum phase transition in Zr isotopes, where a level crossing occurs between spherical and strongly deformed states without mixing \cite{togashi2016}.  
The search for other cases of the second-order quantum phase transition is of extreme interest in clarifying nuclear dynamics.

Certain properties of the critical (transition) point of the second-order phase transition are seen around $N$=66.   
Figure \ref{fig:Tplot}(e) shows the T-plot of the $0^+_1$ state extending over a wide area though not reaching the spherical limit.  This is consistent with a large quantum fluctuation typical for the critical point.  The T-plot circles of the $2^+_1$ state are discretely displaced from those of the $0^+_1$ state, keeping the $2^+_1$ state in the deformed phase.  The angular momentum $J$ can thus be another control parameter.  This $0^+_1$-$2^+_1$ difference causes a suppression of the B(E2) value, to be concrete, due to more neutrons in the 3$s_{1/2}$ (1$h_{11/2}$) orbit for the $0^+_1$ ($2^+_1$) state.   Since some experiments do not show this suppression, the ``Alternative'' set of TBMEs was introduced mainly for obtaining a larger B(E2) value.  
Figure~\ref{fig:Tplot} (j,k,l) exhibit, respectively, T-plots for $^{110,116,122}$Sn obtained from this set.
A notable difference from the ``Present'' set appears only for $^{116}$Sn, and the overall structure evolution remain unchanged.  A consistent feature is seen in Fig.~\ref{fig:BE2} (a), where    
the dip is shifted only to $N=64$ with the ``Alternative" set.    
Thus, the features around the critical point may give certain constraints on particular TBMEs, keeping the present overall picture basically intact.
 
The magnetic moment of the 2$^+_1$ state has been measured recently \cite{allmond2015}, providing a sensible measure of configurations.  The calculated g-factor of $^{112-124}$Sn is, respectively, 0.13, 0.08, 0.02, -0.01, -0.04, -0.05 and -0.07 in an agreement with these data, whereas other theoretical results are for limited nuclei or deviate more \cite{allmond2015}. 

The T-plot of $^{100}$Sn (panel (a)) is similar to the one for $^{132}$Sn (panel (h)), but the circles are spread more outwards, {\it i.e.}, stronger ground-state correlations, certainly because of the $N$=$Z$ nucleus.  
The Gamow-Teller decay of the $^{100}$Sn $0^+_1$ state to the $^{100}$In $1^+_1$ state was measured, giving the largest  B(GT) value.  The T-plots of these states (panels (a,i)) are similar to each other, suggesting a large B(GT).  In fact, the calculated value, B(GT) =9.2 (10.3), with the usual quenching factor 0.70 (0.74), \cite{suzuki2012} (\cite{caurier2005}), shows a good agreement with this recent experimental value 9.1$^{+2.6}_{-3.0}$ \cite{hinke2014}, implying that the amount of the holes in the 1$g_{9/2}$ orbit shown in Fig.~\ref{fig:occup} is appropriate.

 
In summary, we present a unified description of the structure of even-A Sn isotopes for $N$=50-88 owing to state-of-the-art Monte Carlo Shell Model calculations.
The huge bump of the $B(E2; 0^+_1\rightarrow2^+_1)$ value around $N$=60, a decade-long puzzle/challenge, 
is reproduced by activating protons in the 1$g_{9/2}$ orbit.  
The second-order quantum phase transition is shown to occur around $N$=66 from the moderately deformed phase 
to the pairing (seniority) phase, as seen in E2 matrix elements and $S_{2n}$ values.  
The shape evolution in Sn isotopes is linked with the breaking of the $Z$=50 magic number, and this work presents a new picture involving the shell and shape evolutions and the quantum phase transition.
Experimental studies on relevant physical observables, {\it e.g.}, charge radius, are of urgent interest, as well as similar studies on neighboring nuclei.    

\section{Acknowledgements}

We thank Prof. A. Gade for valuable comments on the manuscript.   
This work was supported in part by HPCI Strategic Program (hp150224),
in part by MEXT and JICFuS
and a priority issue (Elucidation of the fundamental laws and
evolution of the universe)
to be tackled by using Post ``K'' Computer (hp160211, hp170230), 
in part by the HPCI system research project (hp170182),
and by CNS-RIKEN joint project for large-scale nuclear structure
calculations.


\end{document}